\documentclass[aps,prb,twocolumn,groupedaddress,floats,showpacs]{revtex4}
\usepackage{latexsym}
\usepackage{dcolumn}
\usepackage[dvips]{graphicx}
\usepackage{amssymb}
\usepackage{graphics}
\usepackage{amsmath}
\usepackage{epsf}

\begin{document}

\author{Shaffique Adam and M. D. Stiles}
\affiliation{Center for Nanoscale Science and Technology, 
National Institute of Standards and Technology, 
Gaithersburg, Maryland 20899-6202, USA}

\title{Temperature Dependence of the Diffusive Conductivity for 
Bilayer Graphene}

\date{\today}
\begin{abstract}
Assuming diffusive carrier transport, and employing an effective
medium theory, we calculate the temperature dependence of 
bilayer graphene conductivity due to Fermi surface broadening as a function of carrier
density.  We find that the temperature dependence
of the conductivity depends strongly on the amount of disorder.  In the regime 
relevant to most experiments, the conductivity is a function of
$T/T^*$, where $T^*$ is the characteristic temperature set by
disorder.  We demonstrate that experimental data taken from various groups
collapse onto a theoretically predicted scaling function.  
\end{abstract}
\pacs{72.80.Vp,73.23.-b,72.80.Ng}
\maketitle

\section{Introduction}

Monolayer and bilayer graphene are distinct electronic materials.  Monolayer 
graphene is a sheet of carbon in a honeycomb lattice that is
one atom thick, while bilayer graphene comprises two such sheets, with
the first lattice $0.3~\rm{nm}$ above the second.  Since the first
transport measurements~\cite{kn:novoselov2005,kn:zhang2005} in 2005, 
we have come a
long way in understanding the basic transport mechanisms of carriers
in these new carbon allotropes.  (For recent reviews, see
Refs.~\onlinecite{kn:neto2009, kn:dassarma2010a}).

A unique feature of both monolayer and bilayer graphene is that the
density of carriers can be tuned continuously by an external gate from
electron-like carriers at positive doping to holes at negative
doping.  The behavior at the crossover depends strongly on the amount
of disorder.  In the absence of any disorder and at zero
temperature, there are no free carriers at precisely zero doping.
However, ballistic
transport through evanescent modes should give rise to a universal minimum
quantum limited conductivity $\sigma_{\rm min}$ in both
monolayer~\cite{kn:katsnelson2006,kn:tworzydlo2006} and bilayer
graphene.\cite{kn:snyman2007,kn:cserti2007,kn:trushin2010}  
The ``ballistic regime'' should hold so long as the disorder-limited mean-free
path is larger than the distance between the
contacts.~\cite{kn:miao2007,kn:danneau2008}  At finite temperature, the 
thermal smearing of
the Fermi surface gives a density $n(T) \sim T^2$ for monolayer
graphene.  For ballistic
transport in these monolayers, the conductivity $\sigma \sim
\sqrt{|n|}$ for large $n$, 
so $\sigma(T) \sim T$.\cite{kn:bolotin2008b,kn:du2008}
In the absence
of disorder, $\sigma(T)$ interpolates from the universal $\sigma_{\rm
  min}$ to the linear in $T$ regime following a function that depends
only on $T/T_{\rm F}$; ($T_{\rm F}$ is the Fermi temperature).
\cite{kn:muller2009}

Most experiments, however, are in the dirty or diffusive limit, which 
is characterized by
a conductivity that is linear in density (i.e. $\sigma
= n e \mu_c$, with a mobility $\mu_c$ that is independent of both
temperature and carrier density~\cite{kn:morozov2008,kn:zhu2009}), and
the existence of a minimum conductivity plateau~\cite{kn:adam2008a}
 in $\sigma(n)$, with
$\sigma_{\rm min}=n_{\rm rms} e \mu_c/\sqrt{3}$.
$n_{\rm rms}$ is the root-mean-square fluctuation in carrier density
induced by the disorder.    
In bilayer graphene, to our knowledge, all experiments 
are in the diffusive limit.

The purpose of the current work is to calculate 
the temperature dependence of the minimum conductivity plateau in
bilayer graphene.  The temperature dependent conductivity of 
diffusive graphene monolayers is understood to depend largely on
phonons,~\cite{kn:chen2008b} but monolayer and bilayer graphene are
distinct electronic materials and phonons are not expected to be
important for bilayer graphene transport at the experimentally relevant 
temperatures.~\cite{kn:footnote1}

\section{Theoretical Model}

An important difference between monolayer and bilayer graphene is the
band structure near the Dirac point.  Monolayer graphene has the
conical band structure and a density of states that vanishes linearly
at the Dirac point.  Bilayer graphene has a constant density of states
close to the Dirac point from a hyperbolic dispersion.
The tight-binding
description for bilayer graphene~\cite{kn:mccann2006b,kn:nilsson2006b} 
results in a
hyperbolic band dispersion 
\begin{equation}
E_{\rm F}(n) = v_{\rm F}^2 m \left[\sqrt{1
    + n/n_0} - 1 \right],
\label{Eq:band}
\end{equation}
 that is completely specified by two
parameters, $v_{\rm F} \approx 1.1~\times 10^{8}~{\rm cm/s}$ and $n_0
= v_{\rm F}^2 m^2/(\hbar^2 \pi) \approx 2.3~\times 10^{12}~{\rm cm}^{-2}$
(where $h = 2 \pi \hbar$ is Planck's constant).  For very
small carrier density $n \ll n_0$, one can approximate bilayer
graphene as having a parabolic dispersion, although most experiments
typically approach carrier densities as large as $5~\times 10^{12}$.
The density of states for bilayer
graphene is 
\begin{equation}
D(E) = \frac{2 m}{\pi \hbar^2} \left[ 1 + \frac{|E|}{ v_{\rm F}^2 m} \right],
\end{equation}
where the parabolic approximation keeps only the first term.

Understanding the temperature dependence of the conductivity minimum
is complicated for two reasons.  First, there is activation of both
electron and hole carriers at finite temperature. Second, the disorder
induces regions of inhomogeneous carrier density (i.e. puddles of
electrons and holes).  Moreover, tuning the carrier density with a
gate changes the ratio between electron-puddles and hole-puddles,
until at very high density there is only a single type of carrier.
The temperature dependence of the conductivity for bilayer graphene
was studied in Ref.~\onlinecite{kn:nilsson2006b} using a coherent
potential approximation.  While this approach better captures the
impurity scattering and electronic screening properties of graphene,
it does not account for the puddle physics which is our main focus.
Reference \onlinecite{kn:zhu2009} modeled the temperature dependence
of the Dirac point conductivity by assuming that the graphene samples
comprised just two big ``puddles'' each with the same number of
carriers. In the appropriate limits, our results agree with these
previous works.  Below we will provide a semi-analytic expression for
the graphene conductivity by averaging over the random distribution of
puddles with different carrier densities.  This result is valid
throughout the crossover from the Dirac point (where fluctuations in
carrier density dominate) to high density (where these fluctuations
are irrelevant), both with and without the thermal activation of
carriers.

\begin{figure}
\bigskip
\epsfxsize=1.0\hsize
\begin{center}
\epsffile{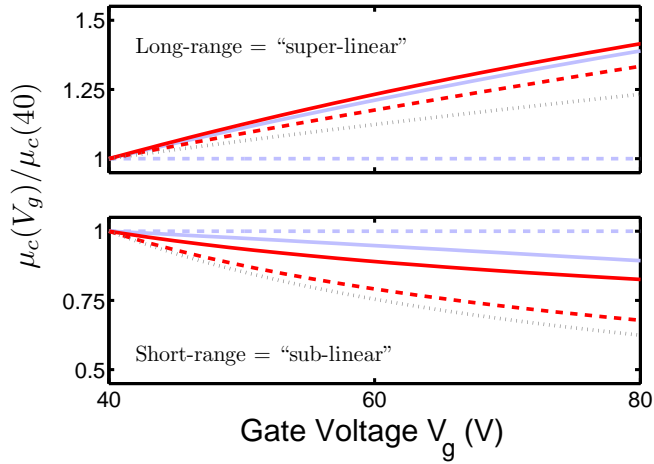}
\end{center}
\caption{\label{fig:mobility} (Color online)
Bilayer graphene mobility as a function of back-gate voltage $V_g$, 
normalized by the mobility at $V_g = 40~{\rm V}$.  Solid lines use
bilayer graphene's hyperbolic dispersion relation, while dashed lines 
are the parabolic approximation valid only for low carrier density.  Upper panel -- 
long-ranged Coulomb impurities.  From bottom to top: 
over-screened (parabolic), RPA (parabolic), Thomas-Fermi (parabolic),
over-screened (hyperbolic), Thomas-Fermi (hyperbolic).  Lower panel: 
short-range (i.e. ``delta-correlated'' or ``white noise'') impurities.
From bottom to top: RPA (parabolic), Thomas-Fermi (parabolic), 
Thomas-Fermi (hyperbolic), unscreened (hyperbolic), unscreened (parabolic).
See Ref.~\onlinecite{kn:dassarma2010a} for definitions of the different
approximations.}
\end{figure}

Given a microscopic model for the disorder, one can compute both
$\mu_c$ and $n_{\rm rms}$.  Shown in Fig.~\ref{fig:mobility} are
results for bilayer graphene mobility assuming both short-range and
Coulomb disorder with different approximations for the screening, and for both
parabolic and hyperbolic dispersion relations.  
As seen from the figure, generically, Coulomb impurities show a
super-linear dependence on carrier density while short-range
scattereres are sub-linear.  Similar to monolayer
graphene,\cite{kn:adam2007a, kn:jang2008,kn:ponomarenko2009} increasing the
dielectric constant tends to decrease (increase) the scattering of
electrons off long (short) range impurities, except in the
over-screened and unscreened limits.  All experiments to date find
the mobility to be linear in gate voltage, so it is unclear what
the dominant scattering mechanism in bilayer graphene is (see also
discussion in Ref.~\onlinecite{kn:xiao2010}).  Further experiments along
the lines of Refs.~\onlinecite{kn:jang2008,kn:ponomarenko2009} are needed.

In what follows we take $\mu_c$ and
$n_{\rm rms}$ to be parameters of the theory that can be determined
directly from experiments: $\mu_c$ can be obtained from low
temperature transport measurements and $n_{\rm rms}$ from local probe
measurements.\cite{kn:deshpande2009b,kn:martin2008,kn:zhang2009,kn:miller2009}
Lacking such microscopic measurements for the samples we compare
with, we treat $n_{\rm rms}$ as a fitting parameter, while taking
$\mu_c$ from experiment.
As a consequence of this parameterization, the results 
reported here do not depend on the microscopic details of the impurity
potential, provided this parameterization reasonably characterizes the
properties of the impurity potential.  
Until more information about the important scattering centers is
determined from experiment, all microscopic models will require a
similar number of parameters such as the concentration of impurities
$n_{\rm imp}$ and their typical distance $d$ from the graphene sheet. 
Further, the results will disagree with experiment unless the choices
give a constant mobility.

A key assumption in this work is the applicability of Effective
Medium Theory (EMT), which describes the bulk conductivity
$\sigma_{\rm EMT}$ of an inhomogeneous medium by the integral
equation~\cite{kn:rossi2008b}
\begin{equation}
\int dn P[n] \frac{\sigma(n) - \sigma_{\rm EMT}}{\sigma(n) + \sigma_{\rm
    EMT}} = 0.
\label{Eq:EqEMT1}
\end{equation}    
$P[n]$ is the probability distribution of the carrier density in the
inhomogeneous medium -- positive (negative) $n$ corresponds to
(electrons) holes, and $\sigma(n)$ is the local conductivity of a
small patch with a homogeneous carrier density $n$.  Ignoring the
denominator, Eq.~\ref{Eq:EqEMT1} gives $\sigma_{\rm EMT}$ equal to
the average conductivity.  The 
denominator weights the integral to cancel the build-up of any
internal electric fields.  The EMT description has been shown to work
well whenever the transport is semiclassical and quantum corrections
and any additional resistance caused by the $p-n$ interfaces between
the electron and hole puddles can be
ignored.\cite{kn:rossi2008b,kn:adam2008d,kn:fogler2008b} It is assumed that the band
structure is not altered by the disorder, which is to be expected for
the experimentally relevant disorder
concentrations.\cite{kn:pershoguba2009} Since we are concerned with
diffusive transport in the dirty limit, we expect that the EMT results
hold for bilayer graphene.

\section{Results}

\begin{figure}
\bigskip
\epsfxsize=1.0\hsize
\begin{center}
\epsffile{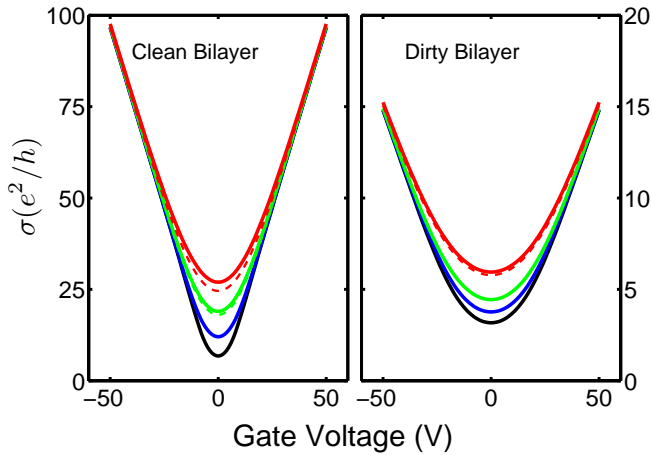}
\end{center}
\caption{\label{fig:main} 
(Color online) Conductivity vs. gate voltage for clean and dirty
  graphene bilayers calculated from Eq.~\ref{Eq:EqEMT1}.  Solid curves
  use the hyperbolic dispersion relation while dashed lines (only
  distinguishable at high temperature) show the parabolic
  approximation.  Choice of parameters were based on experiments of
  Ref.~\onlinecite{kn:morozov2008} (clean) and Ref.~\onlinecite{kn:fuhrer2009}
  (dirty).  Left panel: $\mu_c = 6,750~{\rm cm}^2/{\rm Vs}$, $n_{\rm
    rms} = 4\times 10^{11}~{\rm cm}^{-2}$ and (from bottom to top) T =
  20~{\rm K}, 100~{\rm K}, 180~{\rm K} and 260~{\rm K}. Right panel:
  $\mu_c = 1,100~{\rm cm}^2/{\rm Vs}$, $n_{\rm rms} = 1.25\times
  10^{12}~{\rm cm}^{-2}$ and (from bottom to top) T = 12~{\rm K},
  105~{\rm K}, 171~{\rm K} and 290~{\rm K}.}
\end{figure}

\begin{figure}
\bigskip
\begin{center}
\epsfxsize=1.0\hsize
\epsffile{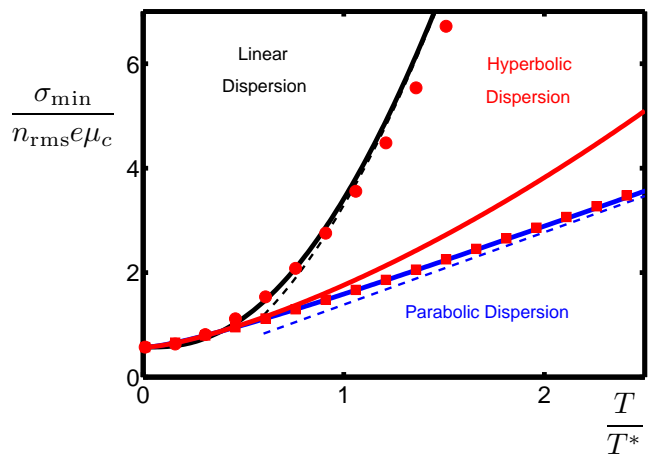}
\end{center}
\caption{\label{fig:scaling} 
(Color online) Minimum conductivity as a function of temperature for
  linear dispersion (upper curve) and parabolic dispersion (lower
  curve) graphene.  Dashed lines show the high temperature asymptotes
  $\sigma_{\rm min} \rightarrow \pi e \mu_c T^2/(3 \hbar^2 v_{\rm
    F}^2)$ for linear and $\sigma_{\rm min} \rightarrow m e \mu_{c} 4
  \ln 2 T/(\pi \hbar^2)$ parabolic cases.  Solid (red) line shows the
  hyperbolic result for $n_{\rm rms} = 10^{12} {\rm cm}^{-2}$.  Also
  shown is that the hyperbolic result extrapolates from the parabolic
theory at large $\alpha = m^2 v_{\rm F}^2 /(\hbar^2 \pi n_{\rm rms})$
becoming similar to the linear dispersion for small $\alpha$.  
(Red squares show results for $\alpha = 100$ and red circles 
are for $\alpha = 0.01$; here we ignore the contribution
from higher bands).}
\end{figure}

To solve Eq.~\ref{Eq:EqEMT1} we make the additional assumption
that the distribution function $P[n, n_g]$ is Gaussian centered at
$n_g$, (i.e. the field effect carrier density
induced by the back gate that is proportional to $V_g$), with width
$n_{\rm rms}$.  (This assumption is justified both
theoretically\cite{kn:morgan1965,kn:stern1974,kn:galitski2007,kn:adam2009} 
and empirically\cite{kn:deshpande2009b}).  Our results are
shown in Fig.~\ref{fig:main}, where as discussed earlier, the
temperature dependence comes from the smearing of the Fermi surface.

At first glance, it is not obvious that the results for clean bilayer
graphene (left panel of Fig.~\ref{fig:main}) and dirty bilayer
graphene (right panel) are closely related.  However, if we
consider scaling the conductivity as ${\tilde \sigma_{\rm EMT}} =
\sigma_{\rm EMT}/(n_{\rm rms} e \mu_c)$, scaling temperature as
$t = T/T^*$, where we define $k_{\rm B} T^* = E_{\rm F}
(n=n_{\rm rms})$, and scaling carrier density as $z =
n/n_{\rm rms}$, we find that for both the linear band dispersion $(n \gg
n_0)$ and the parabolic band dispersion $(n \ll n_0)$, the scaled
functions ${\tilde \sigma_{\rm EMT}}(z,t)$ each follow a
universal curve.  This is illustrated in Fig.~\ref{fig:scaling} where
we show the temperature dependence of the minimum conductivity.  The
results for the hyperbolic dispersion (which is the correct
approximation at experimentally relevant carrier densities),
depends on an additional parameter 
$\alpha = n_0/ n_{\rm rms})$.\cite{kn:footnote2}  
 
The scaling function for the hyperbolic dispersion 
extrapolates from the parabolic theory at large $\alpha$ becoming 
similar to the linear result for small $\alpha$.  For the 
experimentally relevant regime $\alpha \approx 1$ the hyperbolic 
result depends only weakly on $\alpha$ and is indistinguishable
from the parabolic result for $T \lesssim 0.5~T^{*}$.

\begin{figure}
\bigskip
\epsfxsize=1.0\hsize
\begin{center}
\epsffile{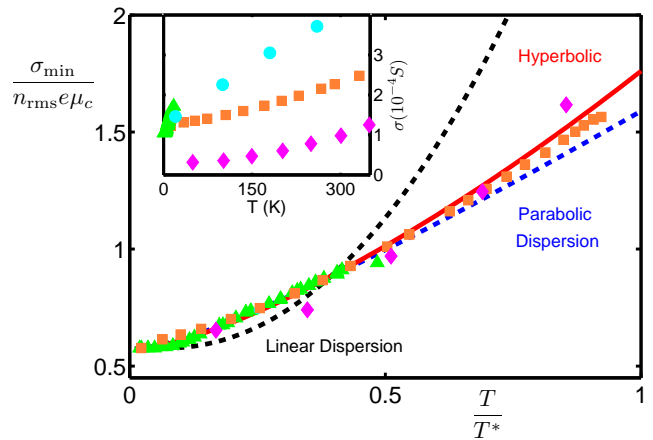}
\end{center}
\caption{\label{fig:expts} 
(Color online) Same results as in Fig.~\protect{\ref{fig:scaling}}
  showing comparison with experimental data from several groups.
  Inset shows the unscaled experimental data, while the main panel
  shows that the data collapses onto the theoretical curve with one
  scaling parameter ($n_{rms}$), where for each of these samples, we
  also use the value of mobility reported by the authors and obtained
  from a separate low temperature measurement.  Green triangles show
  suspended bilayer data from Ref.~\onlinecite{kn:feldman2009} using $\mu_c
  =1.4~\mbox{\rm m}^2/\mbox{Vs}$ and $T^* = 36~{\rm K}$.  Orange
  squares (Ref.~\onlinecite{kn:fuhrer2009}) and diamonds
  (Ref.~\onlinecite{kn:zhu2009}) are bilayers on a SiO$_2$ substrate with
  $\mu_c =0.11~\mbox{\rm m}^2/\mbox{Vs}$, $T^* = 530~{\rm K}$ and
  $\mu_c =0.045~\mbox{\rm m}^2/\mbox{Vs}$ and $T^* = 290~{\rm K}$.
  Cyan circles show the four data points of
  Ref.~\onlinecite{kn:morozov2008}, with $\mu_c =0.675~\mbox{\rm
    m}^2/\mbox{Vs}$, $T^* = 80~{\rm K}$, which are off-scale in the
  main panel.}
\end{figure}

This analysis suggests that $\sigma_{\rm min}(T)/(e \mu_c)$, which can
be taken directly from experiment, is not a function of $\mu_c$, but
only $n_{\rm rms}$.  We take results from a set of experiments in very
different regimes (see the inset of Fig.~\ref{fig:expts}) and choose
$n_{\rm rms}$ to fix the value of $\sigma_{\rm min}(T)/(n_{\rm rms}e
\mu_c)$ at $T=0$.  Then using $k_{\rm B} T^*(n_{\rm rms}) = E_{\rm
  F}(n_{\rm rms})$ to scale the
temperature, all of the results lie on top of the theoretical curve
computed using the hyperbolic dispersion, see Fig.~\ref{fig:expts}.
The theoretical curve with which they agree is distinct from similar
curves calculated for a  linear dispersion and for the purely
parabolic dispersion at high $T/T^*$.  We note
that the scaling function is more complicated than a line.  The
calculation reproduces not only the initial slope as a function of
temperature, but the crossover to higher temperature behavior.  
For the parabolic dispersion, which agrees at low temperatures, the
conductivity 
extrapolates from $\sigma_{\rm min}(T \rightarrow
0)/(n_{\rm rms}e \mu_c) \approx 3^{-1/2}$ at low temperature to
  $\sigma_{\rm min}(t \gg 1)/(n_{\rm rms}e \mu_c) \approx (2 \ln 2) t$ at
  high temperature, with a crossover temperature scale of $T \approx
  T^*/2$.  
In the
future, it should be possible to further test this agreement by
measuring $n_{\rm rms}$
experimentally.~\cite{kn:deshpande2009b,kn:martin2008,kn:zhang2009,kn:miller2009}



\begin{figure}
\bigskip
\epsfxsize=1.0\hsize
\begin{center}
\epsffile{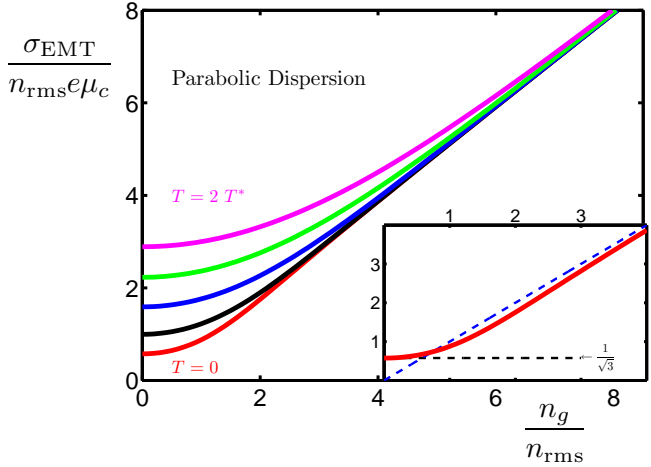}
\end{center}
\caption{\label{fig:parabolic} 
(Color online) Bilayer layer graphene conductivity as a function of
  temperature and carrier density for $T/T^* = 0,0.5,1,1.5$, and $2$.
  Inset shows a close-up of the zero temperature minimum conductivity
  (which is the same for both monolayer and bilayer graphene).  The
  dashed horizontal line shows the result for $\sigma_{\rm min}$, while
  the other dashed line is the high-density transport regime.  The solid line
  (Eq.~\ref{Eq:EMT}) captures the full crossover from the regime where
  the conductivity is dominated by the disorder induced carrier
  density fluctuations, to the semiclassical Boltzmann transport
  regime.}
\end{figure}

One feature of Fig.~\ref{fig:main} and Fig.~\ref{fig:expts} is that
for most of the experimentally relevant regime, the temperature
dependence of the conductivity calculated using the parabolic
approximation provides an adequate solution.  This limit has been
treated in contemporaneous work~\cite{kn:dassarma2010,kn:lv2010}
treating this problem with different approximations and reaching
similar conclusions.  To better understand the emergence 
of a universal
scaling form, we consider the conductivity for a parabolic band dispersion.
Using the scaled variables defined above,
we can manipulate Eq.~(\ref{Eq:EqEMT1}) into the dimensionless form
\begin{eqnarray}
\label{Eq:EMT}
\int_0^\infty dz \exp\left[ -z^2/2 \right]
\cosh \left[ z_g z\right] \frac{H[z,t]
- {\bar \sigma}[z_g, t]}{H[z,t] +
{\bar \sigma}[z_g, t]} = 0,
\end{eqnarray}
where $z_g = n_g/n_{\rm rms}$ and we have written the local
conductivity as $\sigma(n,T) = n_{\rm rms}e \mu_c H(z,t)$.  Below we
calculate the dimensionless function $H(z,t)$ assuming thermally
activated carrier transport with constant $n_{\rm rms}$ and $\mu_c$
and explicitly show that it depends only on scaled variables $z =
n/n_{\rm rms}$ and $t = T/T^*$.  With the analytical results for
$H(z,t)$ discussed below, this implicit equation can be solved either
perturbatively or by numerical integration to give $\sigma_{\rm EMT}$.
The results of this calculation are shown in Fig.~\ref{fig:parabolic}.

To proceed, we calculate the function $H(z,t)$.  For thermal
activation of carriers, the chemical potential $\mu$ is determined by
solving for $n_g = n_e - n_h$,\cite{kn:hwang2008b} where
\begin{eqnarray}
n_e(T) &=& \int_0^\infty dE~D(E) f(E,\mu, k_{\rm B} T), \nonumber \\ 
n_h(T) &=& \int_{-\infty}^0 dE~D(E) \left[1- f(E,\mu, k_{\rm B}
  T)\right],
\end{eqnarray} 
where $f(E, \mu, k_{\rm B} T)$ is the Fermi-Dirac function and $k_{\rm
  B}$ is the Boltzmann constant.  
For $T=0$, only majority carriers are present, while for
$T\rightarrow \infty$, activated carriers of both types are present in
equal number.  Within the parabolic approximation, we find $n_{e(h)} = n_g
(T/T_{\rm F}) \ln\left[1 + \exp(\mp \mu/k_{\rm B} T)\right]$ and $\mu
= E_{\rm F}$. Using $\sigma(n,T) = (n_e+n_h) e \mu_c$, we obtain
\begin{equation}
H(z,t) = z + 2 t \ln\left[1+ e^{-z/t}\right].
\end{equation}  
\noindent This demonstrates that
Eq.~\ref{Eq:EMT} depends only on the scaled variables, guaranteeing
that ${\tilde \sigma}_{\rm EMT}$ is a function only of $T/T^{*}$ 
and $n_g/n_{\rm rms}$ as shown in Fig.~\ref{fig:parabolic}.  

A similar analysis can be done for the hyperbolic dispersion.  We find
\begin{eqnarray}
H(z,t, \alpha) = &&\frac{z}{\xi + 2} \left[4 t g \ln[1 + e^{-y/tg}] 
+ 2y \frac{}{}\right.   \nonumber \\
&& \mbox{} \left. + \frac{(t g \pi)^2\xi}{3} + \xi y^2 \right],
\label{Eq:mu}
\end{eqnarray}
where $g(z,\alpha) = T^*/T_{\rm F}$, $\xi(z,\alpha) = -1 + \sqrt{1+z/\alpha}$,
and the scaled chemical potential $y = \mu/E_{\rm F}$ is given by
\begin{eqnarray}
 y = \frac{1}{2} \left[ 2 + \xi - 2 \xi (t g)^2({\rm Li}_2(-e^{-y/tg})
-  {\rm Li}_2(-e^{+y/tg})) \right],
\label{Eq:ef}
 \end{eqnarray} 
where ${\rm Li}_2(z)= \int_z^0 dt~t^{-1}\ln(1-t)$ is the dilogarithm
function.  Only for $\alpha \gg 1$ and $\alpha \ll 1$ does $H(z,t, \alpha)$
become independent of $\alpha = n_0/n_{\rm rms}$ giving the universal scaling 
forms for linear and parabolic dispersions, respectively.

\section{Conclusion}

In summary, we have developed an effective medium theory that captures
the gate voltage and temperature dependence of the conductivity for
bilayer graphene.  The theory depends on two
parameters: $n_{\rm rms}$ that sets the scale of the disorder, and
$\mu_c$ the carrier mobility.  
These could be computed {\it a priori}
 by assuming a microscopic model for the disorder potential and its
 coupling to the carriers in graphene.  Alternatively, one could use an
empirical approach where one uses experimental data at $T=0$ to
determine the parameters and use the theory to predict the temperature
dependence.  

Our main finding is that experimental data taken from various groups
collapse onto our calculated scaling function where the disorder sets
the scale of the temperature dependence of the conductivity.
This further suggests that even some suspended bilayer samples are
still the the diffusive (rather than ballistic) transport regime.

\acknowledgments

We thank M.\ Fuhrer and K.\ Bolotin for suggesting this problem and for
useful discussions.  SA also acknowledges a National Research Council
(NRC) postdoctoral fellowship. \\


\begin{thebibliography}{43}
\bibitem[{kn:({\natexlab{b}})}]{kn:footnote2}
\bibinfo{journal}{A useful way to think about the temperature dependent
  transport for the hyperbolic band dispersion is that it comprises two
  additive channels, one ``parabolic-like'' that is always present, and one
  ``linear-like'' relevant only at high carrier density or high temperature
  (although this simple picture is somewhat complicated by the fact that the
  chemical potential $\mu$ couples the two channels and needs to be calculated
  self-consistently, see Eq.~\ref{Eq:ef}).}
\end{thebibliography}

\end{document}